\documentclass[12pt,psfig]{article}
\usepackage{graphicx}
\usepackage{bm}
\usepackage{amssymb}
\newcommand{\be}{\begin{equation}}
\newcommand{\ee}{\end{equation}}
\newcommand{\bef}{\begin{figure}}
\newcommand{\eef}{\end{figure}}
\newcommand{\bea}{\begin{eqnarray}}
\newcommand{\eea}{\end{eqnarray}}
\newcommand{\bx}{\boldsymbol{x}}

\newcommand{\bh}{\boldsymbol{h}}
\newcommand{\bs}{\boldsymbol{s}}

\newcommand{\by}{\boldsymbol{y}}
\newcommand{\bF}{\boldsymbol{F}}

\newcommand{\bw}{\boldsymbol{w}}
\newcommand{\bpsi}{\boldsymbol{\psi}}

\begin{document}
\title{Singularities in $d-$dimensional Langevin equations with anisotropic multiplicative noise and lack of self-adjointness in the corresponding Schr\"odinger equation}  
\author{{\bf Andrea Gabrielli$^{*,\dag}$}\\ 
{\small $^{(*)}$Dipartimento Ingegneria Civile, Informatica e delle Tecnologie Aeronautiche, }\\
{\small Universit\`a degli Studi ``Roma Tre", Via Vito Volterra 62, 00146 - Rome, Italy;}\\ 
   {\small $^{(\dag)}$``Enrico Fermi" Research Center and Museum (CREF),} \\
    {\small Via Panisperna 89A, 00184 - Rome, Italy}}

\hyphenation{Lan-ge-vin}
\baselineskip20pt \maketitle
\begin{abstract}
In this paper we analyse $d-$dimensional Langevin equations in \^Ito representation characterised by anisotropic multiplicative noise, composed by the superposition of an isotropic tensorial component and a radial one, and a radial power law drift term. This class of model is relevant in many contexts ranging from vortex stochastic dynamics, passive scalar transport in fully developed turbulence and second order phase transitions from active to absorbing states. The focus of the paper is the system behaviour around the singularity at vanishing distance depending on the model parameters. This can vary from regular boundary to naturally repulsive or attractive ones. The work develops in the following steps: (i) introducing a mapping that disentangle the radial dynamics from the angular one, with the first characterised by additive noise and either logarithmic or power law potential, and the second one being simply a free isotropic Brownian motion on the unitary sphere surface; (ii) applying the Feller's - Van Kampen's classification of singularities for continuous Markov processes; (iii) developing an Exponent Hunter Method to find the small distance scaling behaviour of the solutions; (iv) building a mapping into a well studied Schr\"odinger equation with singular potential in order to build a bridge between Feller's theory of singular boundaries of a continuous Markov process and the problem of self-adjointness of the Hamiltonian operator in quantum theory.
\end{abstract}

\section{Introduction}

Multiplicative noise in stochastic differential equations is a common feature of large variety of stochastic systems ranging from finance \cite{black-scholes}, to passive scalar transport in turbulence \cite{falkovich} , ecology and population biology \cite{ecology}, epidemiology and contact processes \cite{mamunoz3} and other fields. Moreover, diffusive dynamics with anisotropic multiplicative noise are important in different domains ranging from pattern formation \cite{kharchenko} to atmospheric dynamics \cite{sura} and image processing \cite{gao}.
Solving models characterised by this kind of noise and understanding their behaviour in the region where the noise vanishes is therefore of great scientific importance to understanding the dynamics of  many physical and natural systems. 

In the case of one-dimensional continuous Markov processes, i.e. Langevin stochastic differential equations (LSDE), multiplicative noise is not a big issue as thanks to the Lamperti transform \cite{yacine} derived from integration of the the \^Ito formula \cite{gardiner}, it is always possible to {\em reduce} the Langevin equation with multiplicative noise into an equivalent Langevin equation with additive noise and appropriate drift term. This permits to reduce the study the multiplicative system to the analysis of an ordinary Brownian motion in an external potential and then map the results back to the original variables.
In $d>1$ the situation becomes more difficult and {\em reducibility} is not always possible, i.e. an exact map from a Langevin equation with anisotropic multiplicative noise into another one with additive noise not always exists. Necessary and sufficient conditions for reducibility have been found in implicit form \cite{yacine} so that to find the transformation, even when the system is reducible, is not a difficult task unless the multiplicative noise is isotropic.

In this paper we study the case of a $d-$dimensional Langevin equation in \^Ito representation characterised by anisotropic multiplicative noise composed by the superposition of an isotropic tensorial component and a radial one, and a radial power law drift term. For this large class of models, that include, in the case of null drift, the celebrated Kraichnan ensemble that model in terms of Markovian stochastic processes the dynamics of the distance of two particles of a passive scalar in a fully developed turbulent flow \cite{falkovich, vergassola}, we develop a set of tools to completely characterise the system behaviour around the singularity.

We will see that, even though in general it is not possible to map the system in a $d-$dimensional LSDE with isotropic additive noise, it is always possible to introduce a change of variables (Sect.~\ref{mapping}) that decouple the dynamics of the angular components from the radial one so that at each time the motion is composed of a free Brownian motion on the unitary sphere and a radial Bessel process \cite{bessel2}, i.e a one dimensional Langevin equation with with additive noise and logarithmic potential.
The singular behaviour of the model at small distances is therefore completely determined by this last equation. For this equation, rather than focusing on exact solutions with particular asymptotic assumptions \cite{dechant,log-exact}, we develop the complete classification of the {\em natural} or {\em regular} boundary conditions {\em \`a la} Feller (Sect.~\ref{Feller}). Then we proceed to an asymptotic scaling analysis at small distances generalising to our case the Frobenius method of power series expansion which confirms Feller's classification. Finally, we study the relation between the singular behaviour of the studied stochastic processes and the problem of self-adjointness of the corresponding Schr\"odinger operator in quantum mechanics.

In this way we obtain a full characterisation of a large class of $d-$dimensional continuous Markovian stochastic processes with anisotropic multiplicative noise including the Kraichnan ensemble as a very important case.

\section {\^Ito formula for a change of variables}
\label{mapping}

We start by considering the following $d-$dimensional general Markovian stochastic differential equation (SDE) (or
Langevin equation) in \^Ito representation:
\be
\dot{\bx}(t)=\bh[{\bx}(t)]+\hat G[{\bx} (t)]{\bs} (t)\,,
\label{fund-eq}
\ee 
where $\bh({\bx})$ is the drift force vector, $\hat G({\bx})$ is an
${\bx}-$dependent $d\times d$ coupling matrix and ${\bs}(t)$ is a 
delta correlated white noise, i.e. satisfying:
\[\begin{array}{l}
\overline{{\bs}(t)}=0\\
\overline{s_i(t)s_j(t')}=\delta_{ij}\delta(t-t')
\end{array}
\]
If $\bh=0$ and $\hat G({\bx})$ explicitly depends on $\bx$, we have the so-called pure {\em multiplicative noise} case:
\be
\dot{\bx}(t)=\hat G[{\bx} (t)]{\bs} (t)\,.
\label{mult}
\ee
Multiplicative noise finds a wide range of applications in statistical physics ranging from statistical field theory for absorbing phase transitions \cite{mamunoz3, mamunoz2} to turbulence \cite{vergassola}.

Let us suppose to have a function $y({\bx})$ of the position ${\bx}$.  By applying \^Ito formula, it
is possible to show \cite{gardiner} that the \^Ito-Langevin equation
satisfied by $y$ is 
\be 
\dot y(t)=\frac{\partial y}{\partial x_k}h_k
+\frac{1}{2}H_{kl} \frac{\partial^2}{\partial
  x_k\partial x_l}y+ 
\frac{\partial y}{\partial x_k}G_{kl}s_l\,,
\label{y-Langevin}
\ee 
where $\hat H=\hat G\hat G^t$, with $\hat G^t$ the transposed of
$\hat G$, is the covariance matrix of the noise term in Eq.~(\ref{fund-eq})
and we have assumed the Einstein notation for sum over repeated indices as in the rest of the paper.
If $y$ is one of the $d$ components of a
vector ${\by}={\by}({\bx})$, we can write the
\^Ito-Langevin equation for ${\by}$ in a more compact vectorial form 
\be
\dot{{\by}}(t)=\nabla_{{\bx}}{\by}\cdot \bh(\bx)|_{{\bx}={\bx}({\by})}+\frac{1}{2}\{[\hat
  H\nabla_{{\bx}}]\cdot\nabla_{{\bx}}\}{\by}|_{{\bx}={\bx}({\by})}
+\left[\frac{\partial {\by}}{\partial {\bx}}\right][(\hat
  G{\bs})]|_{{\bx}={\bx}({\by})}\,,
\label{y-Langevin2}
\ee 
where with $[\partial {\by}/\partial {\bx}]$ we mean the Jacobian matrix with elements
$[\partial {\by}/\partial {\bx}]_{ij}=\partial y_i/ \partial x_j$ and
we have used that by definition $\hat H=\hat H^t$.

In this paper on the power law radial-isotropic case in which $\bh= K x^\gamma \hat{\bx} $ (i.e. drift parallel to ${\bx}$ with power law amplitude in $x$), and 
the noise covariance matrix, often called {\em diffusion tensor}, is given by
\be
H_{ij}=G_{il}G_{jl}=2x^\xi\left[a\delta_{ij}+b{x_ix_j\over x^2}\right]
\,,
\label{kraich}
\ee 
where $0\le \xi\le 2$ and $a$ and $b$ are such that the matrix $
\hat H\equiv((H_{ij}))$ is positive definite at all $\bx$. This is equivalent to 
the constraints $a>0$ and $(a+b)>0$. 
 
Equation (\ref{kraich}) can be rewritten in the operatorial
form as
\[\hat H=\hat G\hat G^t=2x^\xi(a \hat I+b \hat P_{{\bx}})\,\]
where $\hat I$ is the identity operator and $\hat P_{\bx}$ is the
projection operator along the radial direction $\hat{\bx}$. 
Note that $[\hat P_{{\bx}}]^n=\hat P_{{\bx}}$ for each integer $n\ge 1$.
The coefficient $\chi=1+b/a>0$ provides the degree of anisotropy of the multiplicative noise: (i) for $\chi\to 0^+$ the radial motion gets completely suppressed and the motion reduces to a free Brownian motion on the unitary sphere; (ii) at $\chi=1$ the multiplicative noise becomes perfectly isotropic, and the larger $\chi$ the stronger the multiplicative fluctuations in the radial direction with respect to the angular ones.
Clearly, there are different choices of the matrix $\hat G(\bx)$ satisfying Eq.~(\ref{kraich}). In particular, assuming $\hat G(\bx)= \alpha\hat I+ \beta\hat P_{{\bx}}$ with $\alpha,\beta \in \mathbb{R}$, we have $\alpha=\sqrt{a}$ and $\beta$ satisfying the quadratic equation $\beta^2+2\sqrt{a}\beta=b$.
With this assumptions, the Fokker Planck equation corresponding to the Langevin Eq.~(\ref{fund-eq}) becomes 
\be
{\partial \over \partial t} p(\bx,t)=-K{\partial\over\partial x_i}\left[x^{\gamma-1}x_ip(\bx,t)\right]+\frac{\partial^2}{\partial x_i\partial x_j}\left[x^\xi\left(a\delta_{ij}+b{x_ix_j\over x^2}\right)p(\bx,t)\right]\,.
\label{FP-fund}
\ee
The one-dimensional case can be
recovered by fixing $a>0$ and $b=0$.  

The condition $K=0$ and the choice of the power law with radial anisotropy 
Eq.~(\ref{kraich}) for the noise covariance matrix imply that ${\bx}$ satisfies the pure multiplicative
noise \^Ito-Langevin equation, which at varying parameter $a,b$ and
$\xi$ is known as the {\em Kraichnan ensemble} in the turbulence literature.  This model has been
introduced originally to describe the passive scalar transport by a
highly turbulent flow \cite{kraichnan, falkovich} whose compressibility 
is given by ${\cal P}=[\xi a+(d+\xi-1)b]/(d\xi a+\xi b)$ and can be also obtained by the continuous iteration of spatially correlated stochastic displacement fields \cite{displa, io-cecconi2}. A central problem of the Kraichnan ensemble is to understand under which conditions of compressibility of the fluid different particles of the passive scalar coalesce during the dynamics which belongs to the more general problem of particle coalescence in stochastic dynamics \cite{coha, coha2}. In the fluid dynamics
context th{e ensemble is limited to the values of $a$ and $b$ such that, 
once given $d\ge 2$ and $0<\xi\le 2$, the compressibility is 
$0\le {\cal P}\le 1$. However, the mathematical definiteness of the 
Langevin Equation with multiplicative noise with
(\ref{kraich}) makes acceptable the wider set of pairs $(a,b)$ such that
the covariance of the noise is positive definite:
i.e., $a>0$ and at the same time $(a+b)>0$, which includes as a
particular case the ones satisfying the ``fluid condition'' 
$0\le {\cal P}\le 1$.
The Fokker-Planck equation for the Kraichnan ensemble, corresponding to the Langevin equation defined by (\ref{mult}) and (\ref{kraich}) is
\be
{\partial \over \partial t} p(\bx,t)=\frac{\partial^2}{\partial x_i\partial x_j}\left[x^\xi\left(a\delta_{ij}+b{x_ix_j\over x^2}\right)p(\bx,t)\right]\,.
\label{FP-kraich}
\ee
i.e Eq.~(\ref{FP-fund}) with $K=0$.
In the original Kraichnan ensemble $\bx$ represents the vector distance between two passive scalar particles and the main focus of the model is to study under which conditions on the model parameters $a,b$ and $\xi$, and in particular on the value of the compressibility ${\cal P}$, the two particles spontaneously coalesce.
In the language of continuous Markovian stochastic processes this problem can be rephrased in the study of the singularity in 
$\bx=0$ for which, depending on the values of the model parameters, one can have different behaviors: (i) $\bx=0$ is a regular boundary (i.e. one has to set ``by hand" either repulsive or absorbing boundary conditions to find a defined solution); (ii) $\bx=0$ is {\em natural} repulsive boundary; (iii) $\bx=0$ is a {\em natural} attractive boundary following Feller's theory of boundary conditions of generalized diffusion equations \cite{vankampen, feller}.
This problem has been extensively studied through in \cite{vergassola} and \cite{io-cecconi}.

In order to study the possible behaviors at $\bx=0$ of Eq.~(\ref{fund-eq}) with our assumptions on $\bh$ and $\hat H$ and the corresponding Fokker Planck Eq.~(\ref{FP-fund}) here we propose a different and more illustrative approach. Let us start by analyzing the noise term in Eq.~(\ref{y-Langevin2}). It is
\[\bw(t;{\bx})=\left[\partial {\by}\over \partial {\bx}\right]
[(\hat G{\bs})]|_{{\bx}={\bx}({\by})}\,,\]
i.e.,
\[w_i(t;{\bx})=\frac{\partial y_i}{\partial x_k}G_{kl}s_l\,.\]
In this paper we want to find and study, if it exists, a change of
variables ${\by}={\by}({\bx})$ with both $\bx, \by \in \mathbb{R}^d$
such that {\em in some way} the noise term $\bw$ does not
depend on ${\by}$ (or equivalently on ${\bx}$), so that the
\^Ito-Langevin equation for ${\by}$ is governed by an additive delta
correlated white noise plus a $\by-$dependent drift force.  The problem of mapping Langevin equations with multiplicative noise into equations with additive noise has been treated in different contexts \cite{masoliver, yacine}, but the result are of difficult if not impossible direct explicit applicability in dimensions $d>1$.
It is clear that in general
\[\overline{\bw}=0\,\,\forall\,t,{\bx}.\]
In order to visualize better the problem let us recall the one-dimensional case
with  $0<\xi<2$ and leave the ``smooth" case $\xi=2$ to a following discussion.
In the $1d$ case thanks to the Lamperti transformation \cite{yacine} with $G(x)=\sqrt{2a}x^{\xi/2}$, it is immediate \cite{io-cecconi}
to find that for instance
\[y(x)=\frac{2}{\sqrt{2a}}\frac{x^{(2-\xi)/2}}{2-\xi}\] 
leads to the ordinary Langevin equation for $y$ with additive noise
\be
\dot y(t)=-\frac{\xi}{2(2-\xi)}{1\over y(t)}+s(t)\,.
\label{eq-y-d1}
\ee
Note that by varying $\xi$ from $0^+$ to $2^-$, the ``mass" coefficient $M(\xi)=\frac{\xi}{2(2-\xi)}$ of the
drift $-M(\xi)/y$ term of the above equation increases monotonously from $0^+$
to $\infty$ so that such drift term is always attractive.  As explicitly shown below, this is not the case in higher dimension.  

Let us now analyse the more general case $d\ge 2$ where
we have to take into account the off-diagonal noise terms.  Given the
properties of ${\bs}(t)$, let us now calculate the correlations. It is
simple to find \cite{gardiner} that the covariance matrix of the noise can be written as
\[\overline{\bw(t;{\bx})\bw^t(t';{\bx})}=\left[{\partial {\by}\over
\partial {\bx}}\right]\hat H\left[\partial {\by}\over
\partial {\bx}\right]^t\delta(t-t')\,,\]
that is
\[\overline{w_i(t;{\bx})w_j(t';{\bx})}=\nabla_{{\bx}} y_i
\cdot[\hat H \nabla_{{\bx}} y_j]\delta(t-t')\,.\]
Note that the covariance tensor of the equation of the Langevin equation 
of the variable $\by$
\be
\hat D(\by)=\left[{\partial {\by}\over
\partial {\bx}}\right]\hat H\left[\partial {\by}\over
\partial {\bx}\right]^t\vert_{\bx=\bx(\by)}
\label{D-tensor}
\ee
is nothing else 
the \^Ito diffusion tensor of the diffusion dynamics of the variable
$\by$, i.e., the Fokker Planck equation related to the \^Ito-Langevin 
eq.~(\ref{y-Langevin2}) is
\be
\partial_tp(\by,t)=-\nabla\cdot[{\bf F}(\by)p(\by,t)]+{1\over 2}\nabla\cdot\nabla[\hat D(\by)p(\by,t)]\,,
\label{fokker-y}
\ee
where 
\[{\bf F}(\by)=\nabla_{{\bx}}{\by}\cdot \bh(\bx)|_{{\bx}={\bx}({\by})}+\frac{1}{2}\{[\hat
  H\nabla_{{\bx}}]\cdot\nabla_{{\bx}}\}{\by}|_{{\bx}={\bx}({\by})}\]
is the drift force of the Langevin equation.
Eq. (\ref{fokker-y}) can be rewritten as
\be
\partial_tp(\by,t)=-{\partial\over\partial x_i}[F_i(\by)p(\by,t)]+{1\over 2}{\partial^2\over\partial x_i\partial x_j} [D_{ij}(\by)p(\by,t)]\,,
\label{fokker-y-2}
\ee
where repeated indeces are as usual summed up.
We would like to find ${\by}={\by}({\bx})$ such that the diffusion tensor $\hat D(\by)$ defined in Eq.~(\ref{D-tensor}) 
take a $y-$independent diagonal shape in some appropriate 
local orthogonal coordinates at each point $\by$, so that to describe locally the motion as a simple diffusive motion 
plus an eventual diagonal drift term.
In principle, one should look for a general change of variable $\by=\by(\bx)$.
However, given the isotropic-radial form of the tensor $\hat H(\bx)$, it is reasonable to assume in general 
\[{\by}({\bx})={\bx} f(x)\,,\]
i.e. parallel to ${\bx}$  and with $y(\bx)=xf(x)$.
In this case
\[{\partial y_i\over \partial x_j}=\delta_{ij} f(x)+\frac{x_ix_l}{x}f'(x)\,,\]
where $f'(x)=df/dx$ and which implies that the Jacobian 
$\partial {\by}\over\partial {\bx}$ is symmetric. Hence we can write
\bea
\label{cov}
&\left[{\partial {\by}\over\partial {\bx}}\right]\hat H
\left[\partial {\by}\over\partial {\bx}\right]^t=2
x^\xi\left[f(x)\hat I +xf'(x)\hat P_{{\bx}}\right]
\left[a\hat I+ b\hat P_{{\bx}}\right]
\left[f(x)\hat I +xf'(x)\hat P_{{\bx}}\right]&
\\
&=2x^\xi\left\{af^2(x)\hat I+\left[bf^2(x)+2(a+b)xf(x)f'(x)+
(a+b)x^2[f'(x)]^2)\right]\hat P_{{\bx}}\right\}\,,&\nonumber
\eea
where we have used again that
$\hat P^2_{{\bx}}=\hat P_{{\bx}}$.
It is convenient to rewrite Eq.~(\ref{cov}) by using the local hyperspherical 
coordinates using the fact that in any dimension
\[\hat I=\hat P_{{\bx}}+\sum_{i=1}^{d-1} \hat P_{\psi_i}\,,\]
where $\psi_i$ with $i=1,...,d-1$ is the ortogonal set of hyperspherical 
angles and $\hat P_{\psi_i}$ is the projector along angle $\psi_i$.
Note that $\hat P_{{\bx}}\hat P_{\psi_i}=0$ for any $i=1,...,d-1$ and
$\hat P_{\psi_i}\hat P_{\psi_j}=\delta_{ij}\hat P_{\psi_i}$.
Therefore we can rewrite Eq.~(\ref{cov}) as
\bea
&&\left[{\partial {\by}\over\partial {\bx}}\right]\hat H
\left[\partial {\by}\over\partial {\bx}\right]^t=2
x^\xi\left\{(a+b)\left[f(x)+xf'(x)\right]^2
\hat P_{{\bx}}+af^2(x)\sum_{i=1}^{d-1}\hat P_{\psi_i}\right\}\nonumber\\
&&=2
(a+b)x^\xi\left[f(x)+xf'(x)\right]^2
\hat P_{{\bx}}+D_{\psi}(x)\sum_{i=1}^{d-1}\hat P_{\psi_i}
\label{cov2}
\eea
where $D_{\psi}(x)=2ax^{\xi}f^2(x)$.
As a first step let us impose that the coefficient of $\hat P_{{\bx}}$ be 
independent of $x$ and positive, for instance equal to $1$. This implies:
\be
\left[f(x)+xf'(x)\right]^2=\frac{1}{2(a+b)x^{\xi}}\,.
\label{Px-condition}
\ee
The general solution for $0\le \xi<2$ of this equation is (for $\xi=2$ the solution is different and is treated separately in a subsequent section):
\be
f(x)={k\over x}+\frac{2}{\sqrt{2(a+b)}(2-\xi)}x^{-\xi/2}\,,
\label{gen-sol-f}
\ee
where $k$ is an arbitrary constant. Different values of $k$ correspond to different {\em gauges} for the dependence of the angular diffusion constant on the radial component, even though for any possible gauge the radial motion is in its turn always independent on the angular one.
If we substitute Eq.~(\ref{gen-sol-f}) into the components $\hat D_{\psi_i}$,
proportional to $\hat P_{\psi_i}$, of the diffusion tensor, we get:
\be
 \hat D_{\psi}=2a\left[k^2x^{\xi-2}+\frac{4k}{\sqrt{2(a+b)}(2-\xi)}x^{(\xi-2)/2}+
\frac{2}{(a+b)(2-\xi)^2}\right]
\label{d-psi}
\ee
The simplest and most useful gauge for Eq.~(\ref{d-psi}) is $k=0$ through which the angular diffusion constant becomes independent of the radial coordinate $y$ and permits to obtain the condition of local additive noise, even though in general anisotropic. Other choices of $k$ imply a dependence on $y$ of the angular diffusion constant, i.e. the diffusion constant on the unitary sphere depends on the distance from the origin of axes.

By choosing $k=0$, we get 
\be
\by(\bx)=\bx\frac{2}{\sqrt{2(a+b)}(2-\xi)}x^{-\xi/2}=\hat{\bx}\frac{2}{\sqrt{2(a+b)}(2-\xi)}x^{(2-\xi)/2}\,.
\label{y-x-1}
\ee
Note that not only $\by$ is parallel to $\bx$, but its modulus 
is also an increasing function of $x$, with the singularity $\bx=0$ being
mapped into the point $\by=0$.
With this choice we have that the new \^Ito diffusion tensor can be written as 
\be
\hat D=\hat P_{\by}+\frac{4a}{(a+b)(2-\xi)^2}\sum_{i=1}^{d-1}\hat P_{\psi_i}=
\frac{4a}{(a+b)(2-\xi)^2}\hat I+\left(1-\frac{4a}{(a+b)(2-\xi)^2}\right)\hat
P_{\by}\,,
\label{tensor-y}
\ee 
which is constant and diagonal in the local spherical orthogonal framework and where we have used that $\hat P_{\by}=\hat P_{\bx}$ as $\hat{\by}=\hat{\bx}$. More precisely, it does not depend on the modulus $y$ but only on the local radial direction $\hat{\by}$ through the projection operator $\hat P_{\by}$.

Note that this does not automatically imply that the noise of the new \^Ito-Langevin equation satisfied by $\by$ is a isotropic white noise, i.e. independent of the position and of the orthogonal coordinate system. In fact, in order to get this property, $\hat D$ has to
be the identity matrix or proportional to it.
This is obtained by fixing the following linear condition between $a$ and $b$
\[\frac{a+b}{a}=\frac{4}{(2-\xi)^2}>0\,.\]
In this case indeed $\hat D$ becomes  the identity
$\hat I$ and therefore the noise become white, diagonal and isotropic
with unitary diffusion constant\footnote{It is important to note that for the
  isotropic case $b=0$ (and $d\ge 2$) such a mapping is trivially possible
  only for $\xi=0$, i.e. for the already white noise case in the $\bx$
  variable.}.
  
In order to find the Langevin equation for $\by$ given by
Eq.~(\ref{y-x-1}) we need to find the ``drift term'' 
\[\bF(\by)=\nabla_{{\bx}}{\by}\cdot \bh(\bx)|_{{\bx}={\bx}({\by})}+{1\over 2}\{[\hat
  H\nabla_{{\bx}}]\cdot\nabla_{{\bx}}\}{\by}|_{{\bx}={\bx}({\by})}\] 
following to the 
change of variable given by Eq.(\ref{gen-sol-f}) with $k=0$.
This can be done by writing in general the gradient operator in local spherical 
coordinates
\[\nabla_{\bx}=\hat{\bx} \nabla_x+\sum_{i=1}^{d-1}\hat{\bpsi}_i\nabla_{\psi_i}\,,\]
where $\hat{\bpsi}_i$ is the versor indirection of the angle $\psi_i$ and 
$\nabla_{\psi_i}$ is the component of the gradient along $\hat{\bpsi}_i$, with $\nabla_x={\partial\over\partial x}$ and $\nabla_{\psi_i}={1\over x}g_i(\psi_1,...,\psi_{d-1}){\partial\over\partial \psi_i}$ with appropriate $g_i(\psi_i)$.
By using some tensorial algebra and the following differential rules of the 
local spherical frame versors in any dimension $d\ge 2$
\bea
&& \nabla_x\hat{\bx}=0\nonumber \\
&& \nabla_{\psi_i}\hat{\bx}={1\over x}\hat{\bpsi}_i\,,
\label{tensor1}
\eea 
one can write 
\bea
\bF&=&\hat{\bx}\left\{\frac{K}{\sqrt{2(a+b)}}x^{\gamma-\xi/2}+\frac{x^{(\xi-2)/2}}{2\sqrt{2(a+b)}}\left[2a(d-1)-\xi(a+b)\right] \right\}\nonumber\\
&=&J(a,b,\xi,K)y^{2\gamma-\xi \over 2-\xi}\hat{\by}+C(\xi,d,a,b)\frac{\hat{\by}}{y}\,,
\label{force}
\eea
where 
\[J(a,b,\xi,K)=K\left[{(a+b)(2-\xi)^2\over 2}\right]^{2\gamma-2-\xi\over 4}\]
\[C(\xi,d,a,b)=\frac{2a(d-1)-\xi(a+b)}{2(a+b)(2-\xi)}=\frac{d-1}{\chi(2-\xi)}-\frac{\xi}{2(2-\xi)}\,.\]
Note that $J>0$ for all $K>0$, $0<\xi<2$ and $a$ and $b$ such that $\hat H$ is positive definite (i.e. $\chi>0$).

The one dimensional case is recovered, as expected by putting $d=1$ and $b=0$.
Therefore for the generalised Kraichnan ensemble defined by Eq. (\ref{fund-eq}) and with the isotropic-radial power law assumption for $\bh(\bx)$ and $\hat H(\bx)$ in any dimension $d$, it is always possible to find a change of variable $\by=\by(\bx)$ such that the particle motion can be seen as the composition of a simple isotropic Brownian motion on the unitary sphere (i.e. in the hyperspherical coordinates) centred in the origin with diffusion constant $\frac{4a}{(a+b)(2-\xi)^2}$, and a Brownian motion for the radial variable with additive noise with unitary diffusion constant in a {\em singular} power law force field $F(y)=J\,y^{2\gamma-\xi \over 2-\xi}+Cy^{-1}$, i.e. a potential $V(y)={J\over\alpha}y^{-\alpha}-C\log y$ with $\alpha=2(\xi-1-\gamma)/(2-\xi)$ for $\gamma\ne \xi-1$:
\be
\left\{
\begin{array}{ll}
\dot y(t)=Jy^{2\gamma-\xi \over 2-\xi}+\frac{C}{y}+s_y(t) & \\
\\
\dot\psi_i(t)=\sqrt{\frac{4}{\chi(2-\xi)^2}}s_{\psi_i}(t) & i=1,...,d-1
\end{array}
\right.
\label{system-add}
\ee
where $s_y(t)$ and $s_{\psi_i}(t)$ are independent standard white noises.
The Fokker-Planck equation (FPE) corresponding the first equation of (\ref{system-add}) is
\be
{\partial \over \partial t} p(y,t)=-{\partial \over \partial y}\left[\left(Jy^{2\gamma-\xi \over 2-\xi}+Cy^{-1}\right)p(y,t)\right]+\frac{1}{2}{\partial^2 \over \partial y^2} p(y,t)\mbox{  for }y\ge 0
\label{FP-eq-radial}
\ee
from which it is clear that $y=0$ is a singular boundary where the Fokker-Plank operator ${\cal L}_{FP}=-\partial_y[F(y)\cdot]+\frac{1}{2}\partial_y^2(\cdot)$ looses parabolicity.

The same results for the change of variable $\by=\by(\bx)$ and the Langevin equation (or the equivalent Fokker-Planck equation) satisfied by $y(t)$ can be obtained by rewriting Eq. (\ref{FP-fund}) in $d-$dimensional spherical coordinates, integrating out the angular coordinates to get the $1d$ Fokker-Planck equation for the PDF $q(x,t)$ for radial coordinate $x$ in a similar way to hat done in \cite{io-cecconi}:
\be
{\partial \over \partial t} q(x,t)=(a+b){\partial^2 \over \partial x^2}[x^\xi q(x,t)]-{\partial \over \partial x}[(Kx^{\gamma}+(d-1)ax^{\xi-1} )q(x,t)]\,,
\label{FP-kraich-1d}
\ee
and finally looking for the $1d$ change of variable $y=y(x)$ such that to have a constant and unitary diffusion coefficient.
It is important to note that for $d\ge 2$, such integration over the angular coordinates, generate an additional noise induced drift force $(d-1)a x^{\xi-1}$.

In the case in which $\gamma>\xi-1$ the problem at sufficiently small $y$ (i.e. small $x$) practically reduces to the case of the Kraichnan ensemble obtained for $K=0$, independently of $d$ and $\xi$, for which we have $F(y)=\frac{C}{y}$, i.e. a Brownian motion in a logarithmic potential $V(y)=-C\log y$ \cite{dechant}. In this case, the equivalent Langevin equation of (\ref{system-add}) and the Fokker Planck counterpart are know in the mathematical literature as Bessel process \cite{bessel, bessel2} and finds a lot of applications in physics from stochastic of self-gravitating particles and vortex dynamics \cite{chavanis, bray} to polymer physics \cite{polymer}, biology \cite{dna-log}, diffusive spreading of momenta of two-level atoms in optical lattices \cite{two-level}, single particle models of long ranged interacting systems and beyond \cite{bouchet}.
It is important to note that while in $d=1$ we have $C<0$ for all $\xi\in(0,2)$ and $a>0$, i.e. the related term in  Eq.~(\ref{system-add}) is always attractive, in $d\ge 2$ the sign of $C$ depends on the choice of $a,b$ and $\xi$, i.e. the force can be either attractive or repulsive depending on the choice of the model constants. For $\gamma=\xi-1$ also the first drift term in $J$ is proportional to $1/y$ and we have simply $V(y)=-(J+C)/y$, i.e. the same potential as in the Kraichnan ensemble but with a different coefficient. This case of logarithmic potential $V(y)$ has been studied in different physical contexts by different authors \cite{bray, munoz}. The behaviour around $y=0$ can be qualitatively classified through Feller's classification of singular boundaries \cite{feller}, and quantitatively through the methods developed in \cite{vergassola, io-cecconi}.

In the particular case in which 
$\chi=\frac{4}{(2-\xi)^2}$ the $d-$dimensional additive noise becomes totally isotropic in any orthogonal coordinate frame obtained from the hyperspherical one by rotation, including the Cartesian one. In other words, for this particular choice of the model constants $a$ and $b$ the change of variable $\by=\by(\bx)$ transforms the $d-$dimensional Langevin equation defined by (\ref{mult}) and (\ref{kraich}) with multiplicative noise in a Langevin equation with isotropic additive white noise, i.e. it defines the $d-$dimensional Lamperti transform for this model.

The $J$ term in Eq.~(\ref{system-add}) for $y$ is dominant around the singularity $y=0$ only if $\alpha>1$, i.e. $\gamma<\xi-1$ for which at sufficientluy small $y$ we have a drift derived by the potential $V(y)={J\over\alpha}y^{-\alpha}$. Summarizing, for the behaviour at the singularity $y=0$ we can simply distinguish two simple cases for the potential $V(y)$ around $y=0$: (i) $V(y)=-C\log y$ for $\gamma>\xi-1$, (the situation in which $\gamma=\xi-1$ can be included in this case with the change of the coefficient from $C$ to $J+C$), which corresponds to the case in which the multiplicative noise term in Eq.~(\ref{fund-eq}) is dominant around $x=0$ with respect to the drift; (ii) $V(y)={J\over\alpha}y^{-\alpha}$ with $\alpha=2(\xi-1-\gamma)/(2-\xi)>0$ for $\gamma< \xi-1$ when the drift term in Eq.~(\ref{fund-eq}) substantially determines the behaviour of the Langevin equation around the singularity.

Being the angular components of the dynamics of $\by(t)$ independent and identical simple Brownian motions, the peculiar features of its time-evolution (and therefore of $\bx(t)$) are described by the first equation of the system (\ref{system-add}). In particular it completely determines the behaviour around the singularity at $y=0$ (i.e. $x=0$).

We start from the Feller's classification and then we move to a quantitative assessment of the singularity.

\section{Feller's / Van Kampen's classification of the singularity}
\label{Feller}

Here we present the classification of the singularity $y=0$ (e.g. $x=0$) introduced by Feller \cite{feller, gardiner, peskir, bharucha} in a slightly different form as given by  Van Kampen in Chap. XII of \cite{vankampen} which based on the study for $\epsilon \to 0^+$ of the following three fundamental quantities: 1) exit probability $\pi_\epsilon(y_0, y^*)$ from a point $y=\epsilon$ close to the singular boundary $y=0$ starting from $y_0>\epsilon$ before reaching an arbitrary right boundary $y^*>y_0$; 2) conditional mean first-passage time $\tau_\epsilon(y_0,y^*)$ for $y=\epsilon$; (iii) mean first time passage $\tau_{y^*}(y_0,\epsilon)$ through $y^*$ once a reflecting barrier is put at $y=\epsilon$.

These three quantities can be written as \cite{vankampen}
\bea
&&\pi_\epsilon(y_0, y^*)=\frac{\int_{y_0}^{y^*} dy\,e^{2V(y)}}{\int_{\epsilon}^{y^*} dy\,e^{2V(y)}} 
\label{EP}\\
&&\tau_\epsilon(y_0, y^*)=2\int_\epsilon^{y_0}dy\,e^{2V(y)}\int_{y}^{y^*} dy'\,e^{-2V(y')}
\label{FPT}\\
&&\tau_{y^*}(y_0,\epsilon)=2\int_{y_0}^{y^*}dy\,e^{2V(y)}\int_{\epsilon}^{y} dy'\,e^{-2V(y')}
\label{FPT2}
\eea
Moreover, we note that, putting a reflecting barrier at $y=\epsilon$, the probability $\pi_{y^*}(y_0, \epsilon)$ of reaching an arbitrary point $y^*>y_0>\epsilon$ starting from $y_0$ is always one for each $\epsilon>0$.

\subsection{Case $V(y)=-C\log y$}

Let us start with the case in which at small $y$ we can assume $V(y)=-C\log y$. Exact solutions with {\em ad hoc} assumptions for this case are known \cite{log-exact}, however here we rather focus on a complete classification of the possible behaviours around the singularity. 
\begin{itemize}
\item In this case we see immediately that 
\[\lim_{\epsilon\to 0^+}\pi_\epsilon(y_0; y^*)=0\]
for $C\ge 1/2$, i.e. when $\chi\equiv 1+\frac{b}{a}\le d-1$, i.e. $\frac{d-1}{\chi}\ge 1$, corresponding to a strongly repulsive drift force.  Note that this case is possible only for $d>1$ as $\chi>0$ (see Fig.~\ref{fig:1}). It coincides with the results found in \cite{io-cecconi} and \cite{vergassola} for the case of small compressibility ${\cal P} \le \frac{d-2+\xi}{2\xi}$ of the flow in the Kraichnan ensemble.
This means that the a particle starting at an arbitrary $y_0>0$ has zero probability to reach point $y=0$ independently of how small is $y_0$. 
In the same limit $\tau_\epsilon(y_0, y^*)\to \infty$ and $\tau_{y^*}(y_0,\epsilon)$ stays finite.
For this reason, accordingly to Van Kampen's classification, $y=0$ is called a {\em natural repulsive} boundary and no boundary condition is needed to determine the unique solution of the related Fokker Planck equation (\ref{FP-eq-radial}).
With respect to Van Kampen's classification, Feller's one distinguishes two subcases: (a) $\xi<2$ for which for $y_0=\epsilon\to 0^+$ the mean first passage time $\tau_{y^*}(y_0,\epsilon)$ stays finite and $y=0$ is called and ``entrance" boundary; (b) $\xi=2$ for which in the same limit  $\tau_{y^*}(y_0,\epsilon)\to\infty$ and $y=0$ is, as shown below, a ``proper" natural boundary.
Note that this is the only case possible for any finite degree of anisotropy in the high dimensional limit $d\to\infty$.

\item For $C<1/2$, i.e.  $\chi> d-1$  the probability $\pi_\epsilon(y_0; y^*)>0$ for $\epsilon\to 0$ so that the particle starting at $y_0$ has a finite probability of reaching $y=0$ before any other value $y^*>y_0$. In the same limit for $\xi<2$ also  $\tau_\epsilon(y_0, y^*)$ is finite so that the particle reaches $y=0$ typically in a finite time. As shown below, only for $\xi=2$ we can have  $\pi_\epsilon(y_0; y^*)>0$ and $\tau_\epsilon(y_0, y^*)\to \infty$ for $\epsilon\to 0$ which gives the definition of ``proper" {\em natural attractive} boundary for which there is a finite probability of reaching $y=0$, but the mean time is diverging. Also in this case there is no room for arbitrary boundary condition (e.g. reflecting or absorbing) at $y=0$, but it is {\em  naturally} determined by the properties of the Fokker Planck operator close to the singularity.

For $0<\xi<2$ we distinguish two subcases depending on the value taken by $\tau_{y^*}(y_0,\epsilon)$ for $\epsilon\to 0$.
It is simple to show that for $\tau_{y^*}(y_0,\epsilon\to 0)$ stays finite for $-1/2<C<1/2$, while it diverges for $C\le -1/2$. 
In the first case, that happens for all $\frac{d-1}{\chi}<1$ when $\xi\le 1$ and for $\xi-1<\frac{d-1}{\chi}<1$
when $1<\xi<2$ (corresponding to intermediate compressibility in the Kraichnan ensemble $\frac{d-2+\xi}{2\xi} \le {\cal P} < \frac{d}{\xi^2}$), it means that $y=0$ acts as a regular boundary for which it is necessary to set by hand the reflecting or absorbing (or mixed) boundary condition to determine the unique solution of the Fokker Planck equation (\ref{FP-eq-radial}). The second case, which is present only for $1<\xi<2$ and $\frac{d-1}{\chi}<\xi-1$ (corresponding to large compressibility in the Kraichnan ensemble ${\cal P} \ge \frac{d}{\xi^2}$) instead, one talks of a singular ``adhesive" boundary at $y=0$ \cite{vankampen} (see Fig.~\ref{fig:1}). Indeed while the probability of reaching this boundary $\pi_\epsilon(y_0, y^*)>0$  and the conditional mean first passage time $\tau_\epsilon(y_0, y^*)<+\infty$ for $\epsilon\to 0^+$ independently of $y^*>y_0>\epsilon$, we have  $\tau_{y^*}(y_0,\epsilon)\to\infty$ independently of how close to $\epsilon\to 0$ are $y^*$ and $y_0$. This means that the particle dwells an infinite time in an arbitrarily narrow boundary layer around $y=0$. This narrow layer acts as a ``limbo" state where the particle remains confined and a reflecting boundary condition at $y=0$ it cannot be distinguished from an absorbing one. Therefore there is no room also now for an arbitrary boundary condition, but it is fixed automatically from the diffusion operator close to the singularity. For this reason Feller called it a ``natural exit" boundary. In this case, as in the regular one with absorbing boundary condition, the PDF $p(y,t)$ (and equivalently $p(x,t)$) develops a delta function contribution in $y=0$ with a coefficient growing in time up to absorb all the initial probability measure in this point in the large time limit. It is important to note that this corresponds to the absorbing phase in the field theoretical infinite dimensional case of multiplicative noise \cite{mamunoz2, mamunoz3}.
\end{itemize}

\subsection{The {\em natural} singularities for the case $K=0$ and $\xi=2$}

We now study the case with $K=0$ and $\xi=2$. In this case Eq.~(\ref{cov2}) can be rewritten as
\be
\left[{\partial {\by}\over\partial {\bx}}\right]\hat H
\left[\partial {\by}\over\partial {\bx}\right]^t=2
(a+b)x^2[y'(x)]^2
\hat P_{{\bx}}+2ay^2(x)\sum_{i=1}^{d-1}\hat P_{\psi_i}\,,
\label{D-xi-2}
\ee
where we have again assume a change of variables $\by(\bx)=\hat{\bx} y(x)$, implying $\hat{\by}=\hat{\bx}$. Let us now impose again that the radial tensorial component has unitary diffusion constant:
\[2(a+b)x^2[y'(x)]^2=1\,,\]
whose general solution is 
\[y(x)=\pm\frac{\log(cx)}{\sqrt{2(a+b)}}\,,\]
characterised by the possible double sign and the arbitrary integration constant $c>0$.
By imposing that $y(x)$ is an increasing function of $x$, we fix the sign in the previous equation: $y(x)=\frac{\log(cx)}{\sqrt{2(a+b)}}$. Note that, differently for the case $\xi<2$, now the singularity $x=0$ is mapped into $y\to-\infty$. We have again a gauge to fix for the transformation given by the freedom of the constant $c$ and it is natural to choose $c=1$.
Repeating the tensorial arguments already developed in Sect.~\ref{mapping}
In this way the Langevin equations in the new variables, read
\be
\left\{
\begin{array}{ll}
\dot y(t)=(d-2)a-b+s_y(t) &\\
&\\
\dot\psi_i(t)=\sqrt{a}y \, s_{\psi_i}(t)&i=1,...,d-1\,,
\end{array}
\right.
\label{langevin-xi-2}
\ee
with $y(x)=\frac{\log x}{\sqrt{2(a+b)}}$ and $s_i(t)$ and $s_{\psi_i}(t)$ white, normalised and uncorrelated Gaussian noises.
We see that the Langevin equation for $y$ does depend on the angles $\psi_i$, while, differently from the case $0\le \xi<2$, the equations for the hyperspherical angles depends explicitly on radial distance $x$, i.e. on $y$.

The variable $y(t)$ performs a simple Brownian motion with constant drift.
The associated Fokker Plank equation for $y(t)$ is indeed
\be
\partial_t p(y,t)=[b-(d-2)a]\partial_yp(y,t)+{1\over 2}\partial^2_y p(y,t)
\label{FP-xi2}
\ee
This is an ordinary diffusion equation with constant drift term, i.e. a Brownian motion with a constant drift, whose normalized propagator $p(y,t|y_0,0)$ with initial condition $p(y,0)=\delta(y-y_0)$ takes the simple Gaussian form
\be
p(y,t|y_0,0) = \frac{1}{\sqrt{2\pi t}}\exp \left[-\frac{(y-v_0t)^2}{2t}\right]\,,
\label{prop}
\ee 
where $v_0=(d-2)a-b$. This means that $x$ is log-normally distributed with mean logarithmic value linearly increasing with time, a standard logarithmic deviation increasing as $\sqrt{t}$. It is simple to see that $y$ 
reaches $-\infty$ (i.e. $x$ reaches $0$) with probability one if $v_0<0$ (i.e. $b/a>d-2$) in the large time limit, while it never reaches it for $v_0\ge 0$. 
This can be done more rigorous by applying above Feller's / Van Kampen classification of the boundaries $y\to-\infty$.
In this case Eqs.~(\ref{EP} - \ref{FPT2}) become
\bea
&&\pi_{-M}(y_0, y^*)=\frac{\int_{y_0}^{y^*} dy\,e^{2V(y)}}{\int_{-M}^{y^*} dy\,e^{2V(y)}} 
\label{EP-b}\\
&&\tau_{-M}(y_0, y^*)=2\int_{-M}^{y_0}dy\,e^{2V(y)}\int_{y}^{y^*} dy'\,e^{-2V(y')}
\label{FPT-b}\\
&&\tau_{y^*}(y_0,-M)=2\int_{y_0}^{y^*}dy\,e^{2V(y)}\int_{-M}^{y} dy'\,e^{-2V(y')}
\label{FPT2-b}
\eea
where $V(y)=[b-a(d-2)]y$ and that have to be studied for $M\to\infty$. 
It is simple to see that $0<\pi_{-\infty}<1$ for $b>a(d-2)$, while $\pi_{-\infty}=0$ for $b\le a(d-2)$. At the same time $\lim_{M\to\infty}\tau_{-M}(y_0, y^*)= \infty$ in all cases, while $\lim_{M\to\infty}\tau_{y^*}(y_0, -M)= \infty$ for $b\ge a(d-2)$ and remains finite for $b< a(d-2)$.
This means that $y\to-\infty$, i.e. $x=0$ behaves as a ``proper" natural repulsive boundary for $b\le a(d-2)$ and ``proper" natural attractive one for $b> a(d-2)$
\begin{figure}[ht!]
   \centering
   \includegraphics[width=12 cm]{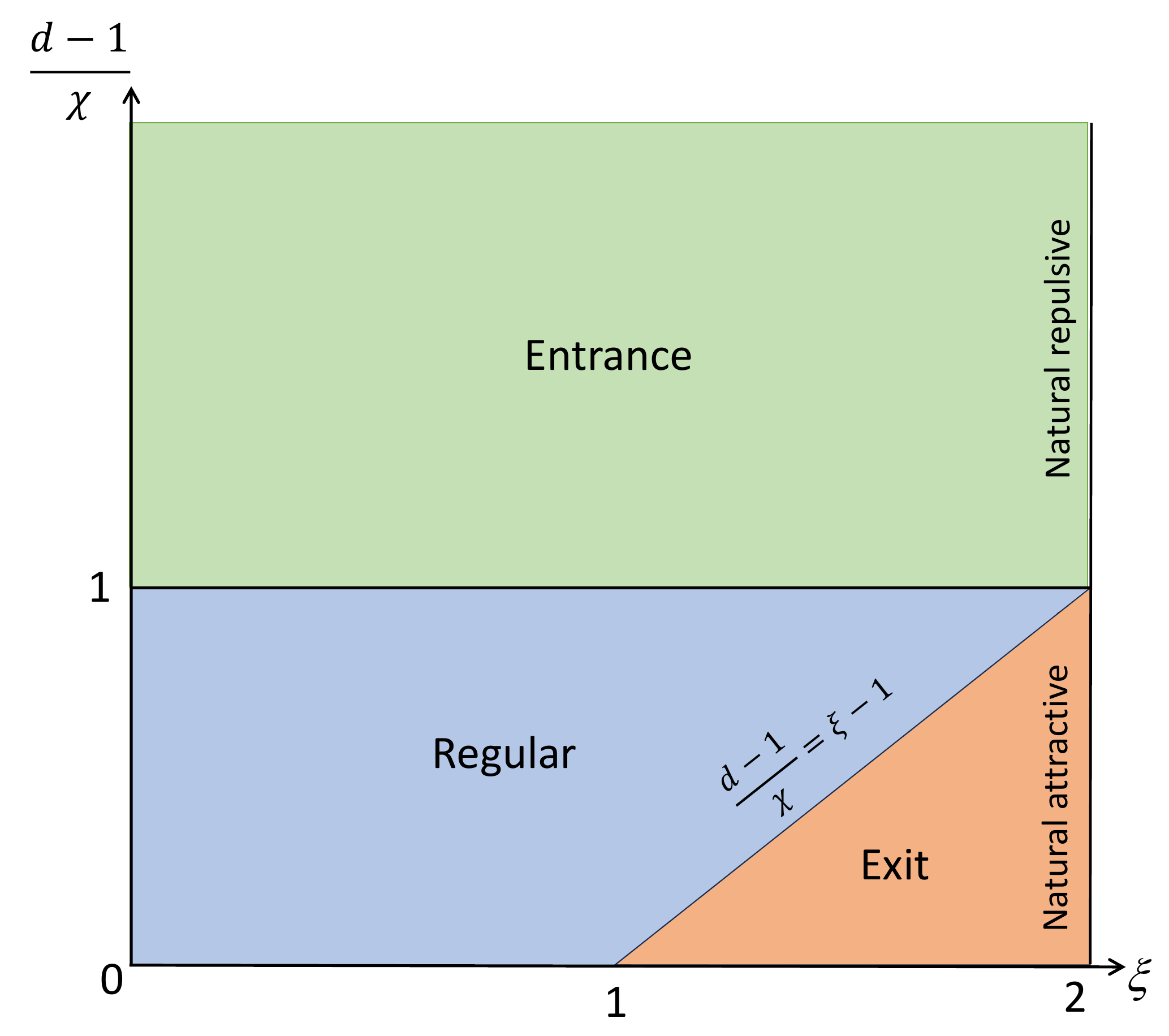}
   \caption{Phase diagram of singularities at $x=0$ for the case $\gamma>\xi-1$, i.e. for $K=0$.}
   \label{fig:1}
\end{figure}

\subsection{Case $V(y)=\frac{J}{\alpha y^\alpha}$ with $\alpha>0$}

The case in which $\alpha=2(\xi-1-\gamma)/(2-\xi)>0$, i.e. $\gamma<\xi-1$, is much simpler to classify as the main distinction is between the repulsive case $J>0$, i.e. $K>0$, and the attractive one $J<0$, i.e. $K<0$.

It is simple to see that in the repulsive case $J>0$ we have $\pi_\epsilon(y_0, y^*)\to 0$ for $\epsilon\to 0$, while $\tau_{y^*}(y_0,\epsilon)$ stays finite for $0<\xi<2$, i.e. $y=0$ is a ``natural repulsive entrance" boundary.
On the contrary, in the attractive case $J<0$,  $\pi_\epsilon(y_0, y^*)>0$ for $\epsilon\to 0$ while the mean first passage time $\tau_\epsilon(y_0, y^*)$ for $y=\epsilon$ stays finite in the same limit and $\tau_{y^*}(y_0,\epsilon)\to \infty$, i.e. $y=0$ is an ``adhesive" or  ``natural exit" boundary similarly to the case $C<-1/2$ above.

\subsection{Considerations on the existence of the stationary state}

Let us assume of setting a reflecting boundary at an arbitrary $y^*>0$. One may ask in which of the above cases the system relaxes towards a well defined stationary state. As explicitly shown in \cite{vankampen}, a stationary state is well defined only in two of the different cases exposed above about the behaviour around $y=0$: (i) $y=0$ is natural entrance boundary, ; (ii) $y=0$ is a regular boundary with imposed reflecting boundary conditions. 
The zero current stationary solution is easily found to be:
\be
p_s(y)=c_0e^{-2V(y)}
\label{stationary}
\ee
with $c_0>0$ a normalisation constant.

In the case in which $V(y)=-C\log y$ we have
\[p_s(y)=c_0y^{2C}\]
and, correspondingly to what above, this is well defined in the neighbourhood of $y=0$ only for $C>-1/2$, where in the repulsive case $C>1/2$ this stationary solution is automatically  attained by the system without need of imposing boundary conditions, while in the intermediate interval $-1/2\le C<1/2$ (regular boundary) it is obtained only by imposing by hand reflecting boundary conditions at $y=0$.

If instead we have $V(y)=\frac{J}{\alpha}y^{-\alpha}$ with $\alpha>0$, we have
\[p_s(y)=c_0\exp\left[-\frac{J}{\alpha}y^{-\alpha}\right]\]
which, in agreement with what above, is well defined only $J>0$ corresponding the ``natural entrance" boundary at $y=0$.

\section{Exponent hunter method (EHM)}
\label{EHM}

As explained above the stationary state exist only in the case of regular boundary at $y=0$ with imposed reflecting boundary condition and in the case of naturale repulsive ``entrance" boundary.
We will now adapt a method introduced in \cite{io-cecconi}, which can be seen as a generalisation to this kind of linear partial differential equations of the power series Frobenius method \cite{frobenius}, to find the small $y$ behaviour of the PDF $p(y,t)$ which satisfies the Fokker Planck equation (\ref{FP-eq-radial}) for the case $\gamma\ge\xi-1$, i.e. $V(y)=-C\log y$ for which the Fokker Planck equation reads:
\be
\partial_t p(y,t)=-C\partial_y\left[p(y,t)\over y\right]+\frac{1}{2}\partial_y^2 p(y,t)\mbox{  for }y\ge 0\,.
\label{FP-eq-radial-log}
\ee
As aforementioned, depending on the value of $C$ we can have different behaviours at $y=0$ from naturally repulsive, to regular to naturally absorbing in which the probability measure progressively accumulates at this singular point through a increasing delta function contribution and there in is no room for arbitrary boundary conditions.

In order to quantitatively characterise these behaviours first we integrate Eq.~(\ref{FP-eq-radial-log}) between $0$ and a generic value $y$.
After some algebra we get
\be
\partial_t P(y,t)=-{C\over y}\partial_y P(y,t)+\frac{1}{2}\partial_y^2 P(y,t)+j(0,t)\,,
\label{FPE-int}
\ee
where $P(y,t)=\int_0^ydy'\,p(y',t)$ and $j(0,t)=\lim_{y\to 0}\left[{C\over y}\partial_y P(y,t)-\frac{1}{2}\partial_y^2 P(y,t)\right]$ is the probability current at time $t$ at $y=0$.

First of all, it is important to notice that Eq.~(\ref{FP-eq-radial-log}) is an homogeneous equation in $y$ and $t$ so that if $p(y,t)$ is any normalised solution, $q(y,t)=\lambda^{1/2}p(\lambda^{1/2}y,\lambda (t+t_0))$ is also a normalised solution with arbitrary $\lambda>0$ and $t_0\in \mathbb{R}$.
Thanks to the homogeneity of Eq.~(\ref{FP-eq-radial-log}), in order to find the small $y$ behaviour of $P(y,t)$ let us assume that it can be written as a power series in $y$:
\be
P(y,t)=\sum_{l=0}^\infty c_l(t) y^{\beta_l}
\label{expansion}
\ee
with $\beta_i<\beta_j$ if $i<j$.
The necessary mathematical conditions that we have to impose is that $P(y,t)$ is a finite, non negative and non decreasing function of $y$ at each $t$.
Putting Eq.~(\ref{expansion}) in (\ref{FPE-int}) we get
\be
\sum_{l=0}^\infty \dot c_l(t) y^{\beta_l}={1\over 2}\sum_{l=0}^\infty \beta_l(\beta_l-1-2C) c_l(t) y^{\beta_l-2}+j(0,t)\,.
\label{eq-series}
\ee
There are just two possibilities in order to match the left hand side member of the above equation with the right one:
\begin{enumerate}
\item $\beta_l=2l$
In this case we can rewrite Eq.~(\ref{eq-series}) as
\[\sum_{l=0}^\infty \dot c_l(t) y^{2l}=\sum_{l=0}^\infty (l+1)(2l+1-2C)c_{l+1}(t) y^{2l}+j(0,t)\,.
\]
From this equation we can derive the differential equations for the coefficients $c_l(t)$:
\be
\left\{
\begin{array}{ll}
\dot c_0(t)=(1-2C)c_1(t)+j(0,t)&\\
&\\
\dot c_l(t)=(l+1)(2l+1-2C)c_{l+1}(t)&\mbox{for  }l\ge1
\end{array}
\right.
\label{eq-coeff}
\ee
Note that these equations are consistent with the above homogeneity condition of the solution which would imply $c_l(t)\sim (t+t_0)^{-l}$, i.e. $c_l(t)y^{\beta_l}\sim [y^2/(t+t_0)]^l$ for $l\ge 1$.

In order to have $P(y,t)\ge 0$ and non decreasing for all possible initial conditions we need to have all the coefficients of the above differential equations non negative, i.e. $C\le 1/2$.  For $C<1/2$ strictly we see that $P(y,t)$ will be dominated at small $y$ by $c_0(t)$ that is a non-negative and increasing term. This means that the PDF $p(y,t)=\partial_y P(y, t)$ develops an increasing delta function contribution $c_0(t)\delta(y)$. As a matter of fact, as shown in the Sect.~\ref{Feller}, for $C\le -/12$ the boundary $y=0$ acts as a natural ``exit"  or absorbing boundary giving rise to increasing delta-function contribution, while for $-1/2<C<1/2$, $y=0$ is a regular boundary at which an absorbing boundary condition can be imposed giving rise to a similar delta-function term. Note that an absorbing boundary condition would require $p(0,t)=0$ and a negative current at the boundary $y=0$. Indeed in term of $P(y,t)$ this means 
\be
j(0,t)=\lim_{y\to 0}\left[{C\over y}\partial_y P(y,t)-\frac{1}{2}\partial_y^2 P(y,t)\right]=c_1(t)(2C-1)<0
\label{current-P}
\ee
and, since $c_i(t)>0$, it is satisfied for $C<1/2$. 
By the same reasoning, we see that $\beta_0=0$ is compatible with reflecting boundary conditions $j(0,t)=0$ only for $C=1/2$ implying $c_0(t)=0$ as, if the initial condition has $c_0(0)=0$ no delta-function develops in time.

\item $\beta_l=2C+1+2l$ and $j(0,t)=0$, implying $\beta_0=2C+1$.
In this case we can rewrite Eq.~(\ref{eq-series}) as
\[\sum_{l=0}^\infty \dot c_l(t) y^{2l}=\sum_{l=0}^\infty (l+1)(2l+3+2C)c_{l+1}(t) y^{2l}+j(0,t)\,.
\]
leading to the following differential system for the coefficient $c_l(t)$
\be
\left\{
\begin{array}{ll}
\dot c_0(t)=(3+2C)c_1(t)+j(0,t)&\\
&\\
\dot c_l(t)=(l+1)(2l+3+2C)c_{l+1}(t)&\mbox{for  }l\ge1
\end{array}
\right.
\label{eq-coeff2}
\ee
By the aforementioned necessary conditions on the function $P(y,t)$, we see that in order to guarantee its finiteness we have to impose $\beta_0=1+2C\ge 0$, i.e. $C\ge -1/2$, which automatically imply that $P(y,t)$ is also non negative and non decreasing as all the coefficients in the system (\ref{eq-coeff2}) are finite and positive.

For $C>-1/2$ strictly we have $P(0,t)=0$ and at the same time from Eq.~(\ref{current-P}) we have $j(0,t)=0$, so that no absorption of probability measure happens at $y=0$ and $y=0$ acts as a reflecting boundary, either naturally for $C\ge 1/2$ when $y=0$ acts as a natural ``entrance" boundary or with imposed reflecting boundary by hand for $-1/2<C<1/2$ when $y=0$ is a regular boundary.
For $C=-1/2$ we have instead $\beta_0=0$ and $j(0,t)=-2c_1(t)<0$ which is incompatible with reflecting boundary condition and an increasing delta function peak at $y=0$ appears, so that $y=0$ acts as a natural ``exit"boundary.

\end{enumerate}


\section{Relation between the case $V(y)=-C\log y$ and the Schr\"odinger equation with singular potential $U(y)\sim y^{-2}$}

In this section we analyze the relationship between the singular behavior of FP equation (\ref{FP-eq-radial-log}) and the problem of self-adjointness of the Hamiltonian of the well studied  $1d$ Schr\"odinger equation with potential $U(y)\sim y^{-2}$.

Let us do some useful preliminary observations on the solution of Eq.~(\ref{FP-eq-radial-log}) in the form of a proper series of eigenfunctions of the Fokker Planck operator ${\cal L}_{FP}=-C\partial_y\left[y^{-1}\cdot\right]+\frac{1}{2}\partial_y^2 [\cdot]$ deriving from the homogeneity in $y$ of such an operator.
Let us assume that $p_\lambda(y)$ is an eigenfunction of the above operator with eigenvalue $\lambda$, i.e
\[-C\partial_y\left[p_{\lambda}(y)\over y\right]+\frac{1}{2}\partial_y^2 p_{\lambda}(y)=\lambda p_{\lambda}(y)\]
It is immediate to realise that, due to the homogeneity on $y$ of ${\cal L}_{FP}$, any function $q(y)=p_{\lambda}(\mu y)$ is also eigenfunction of ${\cal L}_{FP}$ with eigenvalue $\lambda'=\mu^2\lambda$, i.e. $q(y)=p_{\mu^2\lambda}(y)$. This means that the spectrum of ${\cal L}_{FP}$ is infinite and continuous and, as soon as, a single eigenvalue $\lambda_0>0$ appears, the spectrum is no more upper bounded, which is a basic requirement of stability for any Fokker Planck equation. 

The transformation of the above eigenvalue problem in an equivalent Schr\"odinger one, will help us to solve this conundrum shedding light on the mathematical origin of the singular behavior we have above analyzed at $y=0$ and throwing a fundamental bridge between the boundary behaviour theory for diffusion equations of Feller with the Sturm-Liouvile problem in quantum mechanics and in particular with the problem of proper self-adjointness for the Hamiltonian operator in the case of singularities.

In order to map Eq.~(\ref{FP-eq-radial-log}) into a  Schr\"odinger equation let us operate the change of variable $\psi(y,t)=e^{\phi(y)/2}p(y,t)$, with $\phi(y)=-2C\log y$, i.e. $\psi(y,t)=y^{-C}p(y,t)$. Consequently, the non-hermitian FPE transforms in the following hermitian imaginary time Schr\"odinger equation \cite{risken}:
\be
\partial_t \psi(y,t)=\frac{1}{2}\partial_y^2 \psi(y,t)-U(y)\psi(y,t)= {\cal L}_{S}\psi(y,t)\mbox{  for }y\ge 0\,,
\label{Sc-eq-radial}
\ee
where
\[U(y)=\frac{C(C-1)}{2y^2}\]
is the Schr\"odinger potential which is singular in $y=0$ diverging as $y^{-2}$. Being $y=0$ an impenetrable boundary, the standard quantum boundary condition is $\psi(0,t)=0$ with finite derivative. 

The relation between the Schr\"odinger operator and the Fokker Planck one is ${\cal L}_{S}=e^{\phi(x)/2}{\cal L}_{FP}e^{-\phi(x)/2}$.
This is a paradigmatic and well studied ``singular" case of the linear Schr\"odinger equation \cite{reed-simon, essin}. 
Since  ${\cal L}_{S}$ and ${\cal L}_{FP}$ are algebrically similar, they share the same spectrum of eigenvalues $\{\lambda\}$ and the eigenfunctions are related by $\psi_{\lambda}(y)=e^{\phi(y)/2}p_{\lambda}(y)$.
The time independent Schr\"odinger equation can be written as
\be
{d^2 \psi_\lambda(y)\over dy^2}+{k\over y^2}\psi_\lambda(y)=2\lambda\psi_\lambda(y)
\label{sch-eq}
\ee
with $k=C(1-C)$.
As explicitly shown in \cite{essin},  Eq.~(\ref{sch-eq}) has no positive eigenvalue (i.e. no bound states in quantum mechanics where the Hamiltonian operator is ${\cal H}=-{\cal L}_{S}$), not only in the trivial case $k\le 0$, where $U(y)=-{k\over 2y^2}$ provides a quantum repulsive potential lower bounded at null energy, but also in the ``mildly attractive" interval $0<k\le 1/4$. This means that there is no ``energetic" collapse of the system for these values of $k$. On the contrary there are infinite positive eigenvalues for $k\ge 1/4$ (i.e. a continuous set of quantum bounded states up to negatively diverging energy).  It is important to note that $k=1/4$ corresponds to the double value $C=1/2$.
The reason of this strange behavior is that ${\cal L}_{S}$, due to the too fast singularity at $y=0$, even though symmetric, is not a self-adjoint operator in the domain of $L^2$ functions with ``ordinary" quantum boundary conditions (see \cite{essin} and reference therein for the mathematical discussion). 

On the other hand, we have $k\le 1/4$ for all real value of $C$, as from $C(1-C)=k$ we get
\be
C=\frac{1}{2}\pm\sqrt{\frac{1}{4}-k}\,.
\label{C-k}
\ee
This means that for each value of the interaction coefficient $k\le 1/4$ we have two possible values of $C$ symmetric with respect to the value $1/2$ corresponding to the same Schr\"odinger problem. In other words, fixed a value of $C$ in Eq..~(\ref{FP-eq-radial-log}) the same Schr\"odinger equation is obtained also by $C'=1-C$.
This can be simply shown by noticing that in this case the second order linear differential operator $[{d^2 \over dy^2}+k]\cdot$ can be written as the product of two first order operators as follows:
\[\left[{d^2 \over dy^2}+{k\over y^2}\right]\cdot=\left[{d \over dy}+{C\over y}\right]\cdot\left[{d \over dy}-{C\over y}\right]\cdot\,.\]
which implies \cite{essin} that on any possible state $\psi(y)$ the mean value of the energy $\left<\psi |{\cal H}| \psi\right>\ge 0$.

Therefore all the continuous Markov processes (i.e. Bessel processes) defined by Eq.~(\ref{FP-eq-radial-log}) at different values of $C$ (i.e. $a,b,d,\xi$) can be mapped into a Schr\"odinger equation (\ref{Sc-eq-radial}) with the related eigenvalue problem for $k\le 1/4$ with upper bounded spectrum of the operator ${\cal L}_S$ (i.e. lower bounded energies).

The small $y$ behaviour of the eigenfunctions $\psi_\lambda(y)$ can be found through the ordinary Frobenius method of power series expansion \cite{essin}. From Eq.~(\ref{C-k}) it is simple ti show that, for each value of $C$, one can write
\be
\psi_\lambda(y) = y^s\sum_{l=0}^\infty a_l y^{2l}
\label{small-y-psi}
\ee
with two possible values of the exponent $s$: $s_1=C$ and $s_2=1-C$ and the coefficients $a_l$ satisfying the following recursive equations:
\[a_{l}=\frac{C(1-C)}{2l(2l+s-1)}a_{l-1}\mbox{  for }l=1,2,...\]
with $a_0>$ fixed by arbitrary normalisation.

Considering that the correspondence between the eigenfunctions $p_{\lambda}(y)$ of ${\cal L}_{FP}$ and those of ${\cal L}_S$ is simply $p_{\lambda}(y)=y^C\psi_{\lambda}(y)$, the two cases exactly correspond to the classification given in Sects.~\ref{Feller} and \ref{EHM}.
Indeed: (i) for $s=C$ we get 
\[p_\lambda(y) = y^{2C}\sum_{l=0}^\infty a_l y^{2l}\,,\]
which generate a well defined PDF only if integrable at $y=0$, i.e. $C>-1/2$. It is immediate to show that $\lim_{y\to 0}j(y,t)=0$, i.e. the case $s=C$ corresponds either to regular boundary with imposed reflecting boundary condition ($-1/2<C\le1/2$) or to natural repulsive ``entrance" boundary ($C>1/2$).
(ii) On the contrary for $s=1-C$ we have
\[p_\lambda(y) = \sum_{l=0}^\infty a_l y^{2l+1}\,,\]
which generate PDF $p(y,t)$ such that $p(0,t)=0$, i.e.  either $y=0$ as a regular boundary with imposed absorbing boundary condition ($-1/2<C<1/2$) or a natural attractive ``exit" boundary ($C\le-1/2$). In both cases $p(y,t)$ will develop an increasing delta function contribution in $y=0$.
However in order to generate a well defined PDF (non negative and integrable) it is simple to show that the requirement is exactly the one given in Sect.  \ref{EHM}, i.e. $C<1/2$.
Therefore we recover exactly what written above.

Finally, it is important to note that form Eq.~(\ref{small-y-psi}) and $s=C,\,1-C$ for all real values of $C$ at least one of the two independent eigenfunctions $\psi_\lambda(y)$ do not satisfy the ordinary quantum boundary conditions showing a diverging quantum probability current at $y=0$. This is a manifestation of the non self-adjointness of the operator ${\cal L}_S$ in its natural domain \cite{self-adjointness}: the space of $L^2$ functions with the above boundary conditions.
Indeed it is possible to show that, if $\psi$ and $\chi$ are two functions belonging to $L^2(0,\infty)$ such that $\psi(0)=\chi(0)=0$ and finite first derivative, in general we have 
\[\left<{\cal H}\psi|\chi\right>-\left<\psi |{\cal H}\chi\right>\ne 0\]
thanks to surface terms at $y=0$ due to the strong singularity in $y=0$ of the potential $U(y)$. In order to make ${\cal H}$ self-adjoint, more restrictive requirements on the wave functions must be adopted \cite{essin, self-adjointness}.

This shows the strict relationship between Feller's theory of boundary behaviour of continuous Markov processes at singular boundaries and the problem of self-adjointness of the Hamiltonian when the potential presents a strong singularity at the boundary.

\section{Conclusions}

In this paper we have studied the singular behaviour of a large class of $d-$dimensional Langevin equations characterised by anisotropic multiplicative noise with a possible additive radial power law drift term. This set of stochastic systems includes as a particular case the SDEs of the Kraichnan ensemble introduced to describe the possible coalescence of passive scalar particles transported by a fully developed turbulent flows. We focused on the small scale singularity for which we have adopted different theoretical approaches in order to classify and determine the dynamical behaviour. First, we have faced the problem of generalising the one-dimensional Lamperti transform to find a vector variable transformation so that the original SDE with multiplicative and anisotropic noise can be mapped into an equivalent SDE with additive noise and a suitable drift term. We have seen that a suitable variable transformation always exists such that, in local hyperspherical coordinates, one can separate the dynamics of the angular components, coinciding with an isotropic free Brownian motion on the $(d-1)-$dimensional unitary sphere, from the dynamics along the radial direction characterized by a Langevin equation with additive noise (but in general with a different amplitude with respect to the angular components) and a drift term always inversely proportional to the radial coordinate (i.e. deriving from a logarithmic potential) whose coefficient depends on both the exponent and the anisotropy degree of the original multiplicative noise. Thanks to this change of variables the problem of the classification of the possible singular behaviours at short distances reduces to the study of the $1d$ radial Langevin equation in the new variables. We have carried on such a study into the following different steps: (i) we have applied the Feller's - Van Kampen's classification of singular boundaries based on the first passage probabilities and mean times; (ii) then, only for the pure anisotropic multiplicative noise case, we have introduced an {\em exponent hunter} method for the small scale scaling behavior generalizing the Frobenius method for ordinary differential equations. Finally, we have introduced the imaginary time Schr\"odinger equation corresponding to the Fokker Planck equation for the new radial variable in order to relate its singular behavior at short distances to the consequences of the lack of self-adjointness of the Hamiltonian of the Schr\"odinger equation
due to the fast diverging singularity of the potential energy at short distances.

Through these different approaches we are able to provide a full classification and description of the singular behavior of a large class of $d-$dimensional and anisotropic continuous Markovian stochastic processes, with many multi-disciplinary applications, around points at which either the noise vanishes or the drift diverges.

\end{document}